\documentstyle[aps]{revtex}

\begin{document}
\title{$W_{\infty}$-Covariance of
the Weyl-Wigner-Groenewold-Moyal Quantization}
\author{T. Dereli}
\address{ Department of Physics\\
Middle East Technical University\\ 06531 Ankara, Turkey
\\{\footnotesize E.mail: tekin@dereli.physics.metu.edu.tr}}
\author{A. Ver\c{c}in} 
\address{ Department of Physics \\
 Faculty of Sciences, Ankara University\\
06100 Ankara, Turkey
\\ {\footnotesize E.mail: vercin@science.ankara.edu.tr}}
\maketitle

\begin{abstract}
{\small The differential structure of operator bases used in various
forms of the Weyl-Wigner-Groenewold-Moyal (WWGM) quantization
is analyzed and a
derivative-based approach, alternative to the conventional
integral-based one is developed. Thus the fundamental quantum relations
follow in a simpler and unified manner.
An explicit formula for the ordered products of the Heisenberg-Weyl
algebra is obtained.
The $W_{\infty}$ -covariance
of the WWGM-quantization in its most general form
is established. It is shown that the group action
of  $W_{\infty}$ that is realized in the classical 
phase space induces on 
bases operators in the corresponding
Hilbert space a similarity transformation generated by the
corresponding quantum $W_{\infty}$ which provides a projective
representation of the former $W_{\infty}$. Explicit
expressions for the algebra generators in the classical phase space 
and in the Hilbert space are given. It is made 
manifest that this $W_{\infty}$-covariance of the 
WWGM-quantization is a 
genuine property of the operator bases.}
\end{abstract}

\section{INTRODUCTION}
The Weyl-Wigner-Groenewold-Moyal (WWGM) quantization \cite{Weyl} that is 
usually called the phase space formulation of quantum mechanics
has gained a wide popularity  
in many different areas
of physics including statistical mechanics \cite{Kubo},
quantum optics \cite{Klauder}, collission
theory \cite{Carruthers} , and classically chaotic nonlinear
systems \cite{Takahashi,Lee}.
This quantization scheme can be simply stated as an association between 
classical observables (c
-number functions
defined on a classical phase space) and quantum observables (operators 
acting in the corresponding Hilbert space $\cal{H}$). In the mathematical
literature   it is developed
as the theory of pseudodifferential operators  where
the c-number functions determined by the WWGM-quantization are
referred as the symbols of the corresponding Hilbert space operators
 \cite{Folland}.
In fact all the existing methods of quantization 
can be seen as association processes
obeying certain rules \cite{Kirillov}. Therefore, the search
for  possible covariances and hence the invariance properties
that these associations may posses are of fundamental importance. That
is to say, when a member of the associated pair is transformed, determination
of the transformation rule of the other one in a well defined manner
must be the first step of a systematic investigation. In the context
of WWGM-quantization it is unfortunate that only a very restricted class
of covariance properties are specified so far. The main goal of
the present paper is to uncover the covariance properties of the
WWGM-quantization in its as general  form as possible.

The WWGM-quantization associates  the usual product of two operators
$\hat{F_{1}}\hat{F_{2}}$, not with the usual commutative 
product of functions $f_{1}f_{2}$ , but rather with an associative star
product $f_{1} \star f_{2}$ that is in general 
noncommutative, where $f_{1}$ and $f_{2}$ are the c-number 
functions corresponding under the WWGM-quantization to 
the operators $\hat{F_{1}}$ and $\hat{F_{2}}$, respectively. 
Henceforth 
operators and functions of operators will be denoted by 
hat $\hat{}$ over letters. Associativity of the $\star$ product
is inherited from the associativity of usual product of 
the operators. This in turn means that it is not the
Poisson brackets (PB), but its unique $\hbar$ (Planck's constant)
deformation \cite{Bayen}
\begin{eqnarray}
(i\hbar)^{-1}\{f_{1},f_{2}\}_{MB}&=&(i\hbar)^{-1}(f_{1} \star f_{2} -
f_{2} \star f_{1})\nonumber\\
&=&\{f_{1} , f_{2}\}_{PB} + O(\hbar)
\end{eqnarray}
called the Moyal brackets (MB) 
that corresponds to the Lie bracket of operators under the
WWGM-association. Due to the associativity of the $\star$ product
MB obeys  Jacobi identity. Hence, WWGM-quantization sets up
a Lie algebra isomorphism between the Lie algebra of quantum
observables and the resulting Lie algebra of classical observables 
with respect to MB. More precisely, the association depends
on a certain rule of ordering of functions of noncommuting 
operators \cite{Cahill,Agarwal}. For this reason
the star product and therefore
the corresponding MB must be labeled with a parameter specifying the
chosen rule of ordering. 
The $O(\hbar)$ terms in Eq.(1) that are
called "quantum corrections" can be computed to any
desired order of $\hbar$ within the classical regime. 
Hence, the quantization itself can be understood 
as a deformation of classical observables without any
need for introducing a Hilbert space on which the operators act. 
This leads to the fact
that one can define a "pseudomechanics" which has the star product
and MB as its principle ingredients which reduces to  
classical mechanics in the
limit $\hbar \rightarrow 0$ \cite{Bayen,Dunne}. In this case
the dynamics of a classical observable $f$ associated with a
classical Hamiltonian system described by the Hamilton function
$H$ is governed by the "equation of motion"
\begin{eqnarray}
\frac{df}{dt}= (i \hbar)^{-1}\{H,f\}_{MB}.
\end{eqnarray}
The WWGM- quantization 
has a "simple covariance" with respect to affine canonical (i.e.,
inhomogeneous  symplectic) 
transformations \cite{Littlejohn,Han,Osborn}. This affine
canonical covariance  follows from the structure of the automorphism
group of the Heisenberg-Weyl (HW) group $W_{1}$ and
the Stone-von Neumann
theorem \cite{Folland} which in essence states that, upto
a central element generated
by the unit element of the HW-algebra every irreducible representation 
of   $W_{1}$ is unitarily equivalent to the Schr\"{o}dinger
representation
$\hat{D}$ (given by Eq.(3) below). On the other hand, in addition to
the inner automorphisms, automorphism
group of  $W_{1}$
contains the inhomogeneous symplectic group $ISp(2)$ which
is the semidirect product
of the translation group and the symplectic group Sp(2). Thus, we can combine
$\hat{D}$ with an element $\phi \in ISp(2)$ to obtain another representation
$\hat{D}\circ \phi$ which is unitarily equivalent to the Schr\"{o}dinger
representation. This unitary equivalence  provides a double-valued 
metaplectic representation of  ISp(2).  
The affine canonical covariance of the WWGM-quantization
is a simple consequence of this fact.

To the best of our knowledge, the only known covariance of
the WWGM-quantization is the above mentioned metaplectic covariance.
Even this is investigated only for
particular cases of  WWGM-quantization. In the following we prove 
that this quantization scheme has an infinity of covariances
described by the recently found $W_{\infty}$ algebra and $W_{\infty}$
group \cite{Pope}. We wish to warn that there are two $W_{\infty}$'s here. 
The first one
describes the Lie algebra of deformed
classical canonical diffeomorphisms and it is explicitly realized
in the tangent space of the phase space. The other one that acts
in the corresponding Hilbert space  with the usual commutator is the
Lie algebra of  ordered monomials of the HW-algebra generators.
Let us call these the classical $W_{\infty}$ and quantum $W_{\infty}$,
respectively.
Just as it is in the case of the metaplectic covariance, it should be 
emphasized that this
is also a direct consequence of the $W_{\infty}$-covariance of the operator 
bases that are parametrized Schr\"{o}dinger
displacement operators $\hat{D}$
and their Fourier transforms (see Eqs.(4)-(6) below).

We consider only systems with one degree
of freedom for the sake of simplicity. However,  the structure of the
underlying algebra (HW-algebra)  
allows a straightforward generalization to systems with finite or denumerably
infinite number of degrees of freedom. 
Our approach here is distinct from the conventional one
in that differential structure of the bases are given
primary status and are investigated in detail. 
As an alternative to the integral-based
conventional approach we suggest to call this approach
the derivative-based approach.
Except for those that occur in the definitions integrals 
rarely occur in our investigation. Wherever they do occur
integrals should be understood
as double integrations over the whole phase space that is topologically 
equivalent to ${\bf R}^{2}$.

The organization of the paper is as follows. In Section 2 a 
review of the WWGM-quantization
and the definition of  ordered products are given. 
This section fixes the notation and includes formulas
and definitions needed for the subsequent analyses. 
In Section 3 the differential structure of the Weyl basis is obtained
and an explicit formula for
the ordered products is developed. Parametrized Bopp operators are introduced.
Section 4 contains the differential structure
of the generalized Wigner basis and the corresponding Bopp operators.
The announced $W_{\infty}$-covariance of the Weyl and
 Wigner bases are explicitly  established in Section 5.
Quantum deformation of the
canonical diffeomorphisms is also found and the generalized
star product and Moyal brackets are obtained. 
The final
Section 6 contains a  summary of results.
\section{THE WWGM QUANTIZATION AND ORDERED PRODUCTS}
Let us consider the Schr\"{o}dinger representation
of the HW-algebra: $[\hat{q},\hat{p}]=i \hbar \hat{I}$,
where $\hat{q}$
and $\hat{p}$ are the hermitian position and momentum 
operators, respectively. In terms of boson annihilation ($\hat{a}$)
and creation ($\hat{a}^{\dagger}$) operators defined by
$\hat{a}=(a_{0}\hbar \sqrt2 )^{-1}(\hbar \hat{q}+ia_{0}^{2}\hat{p})$,
the defining commutation relation is $[\hat{a},\hat{a^{\dagger}}]=\hat{I}$.
$\dagger$ stands for the Hermitian conjugation and
in terms of a frequency $\omega$ and mass $m$ a length
constant $a_{0}=(\hbar/m\omega)^{1/2}$ is used. 
The identity operator $\hat{I}$ of the algebra generates a $U(1)$ group
that is the center of $W_{1}$. Then  the so called 
displacement operators
\begin{eqnarray} 
\hat{D}(\xi,\eta) = expi(\xi \hat{q}+\eta \hat{p}) \qquad ;
\qquad  \hat{D}(z,\bar{z}) = exp(z\hat{a}^\dagger - \bar{z}\hat{a})
\end{eqnarray}
which act irreducibly in $\cal {H}$ are the representatives
of the coset space $W_{1}/U(1)$
in the real $(\xi,\eta)$ and complex $(z,\bar{z})$ 
parametrization of the group space $W_{1}$, respectively.
$\bar{z}$ denotes the complex conjugation of
$z=-(a_{0}\sqrt2)^{-1}(\hbar \eta - ia_{0}^{2}\xi)$.
The displacement operators, or their suitable parametrizations
form complete  operator bases, in the sense that, any operator
obeying certain conditions can be expanded in terms of
them \cite{Cahill,Agarwal} .
Each  basis is closely connected with
the ordering of noncommuting $\hat{q}$ and $\hat{p}$ (or
$\hat{a}$ and $\hat{a}^{\dagger}$) in the  expansion of operators.
Therefore, not only the  symbols thus obtained but also the
resulting phase spaces are distinct. In essence, the WWGM-quantization
comes into play by considering the parameters of the group space
as the coordinate functions of a phase space. In this sense, the
basis elements play a dual role. On the one hand they are operators
parametrized by the coordinate functions of a 
phase space and acting in $\cal {H}$, and on the other hand they behave as
operator-valued c-number functions defined on the same phase space.

A unified approach to different quantization rules is
achieved by using s-parametrized $(s \in \bf {C})$
displacement operators \cite{Cahill,Balazs}
\begin{eqnarray}
\hat{D}(s)=e^{-i\hbar s \xi \eta /2 }\hat{D}(\xi,\eta) \qquad ;
\qquad \hat{D}(z,s)=e^{s|z|^{2}/2}\hat{D}(z,\bar{z})
\end{eqnarray}
and their Fourier transforms
\begin{eqnarray}
\hat{\Delta}_{qp}(s)&=&(\hbar / 2 \pi)\int\int
e^{-i(\xi q + \eta p)}\hat{D}(s)d\xi d \eta\\
\hat{\Delta}_{Z}(s)&=&(\pi)^{-1}\int\int
e^{-(z \bar{Z} - \bar{z} Z)/a_{0}\sqrt2}\hat{D}(z,s)d^{2}z
\end{eqnarray}
where, $z_{R}$ and $z_{I}$ being the real and the imaginary parts of $z$,
$d^{2}z=dz_{R}dz_{I}$ and  $Z=q+i(a_{0}^{2}/\hbar)p$, $\bar{Z}$ are 
the complex coordinates for the $(q,p)$ phase space. 
The $\hat{\Delta}(s)$ operators are called the s-parametrized 
displaced parity operators. 

It is easy to verify the following trace and unitarity properties
\begin{eqnarray}
\hat{D}^{\dagger}(s)&=&\hat{D}^{\dagger}(0)e^{i\hbar\xi\eta\bar{s}/2}=
\hat{D}^{-1}(\bar{s})\\
\hat{D}^{\dagger}(z,s)&=&\hat{D}^{\dagger}(z) e^{\bar{s}|z|^{2}/2}=
\hat{D}^{-1}(z,-\bar{s})\\
Tr[\hat{D}(s)]&=&(2 \pi / \hbar) \delta(\xi)\delta(\eta)\\
Tr[\hat{D}(z,s)]&=& \pi \delta^{2}(z)
\end{eqnarray}
Another important property, which is independent of $s$,
is the so called displacement property
\begin{eqnarray}
\hat{D}(s)\hat{f}(\hat{q},\hat{p})\hat{D}^{-1}(s)&=&
\hat{f}(\hat{q}+\hbar\eta,\hat{p}-\hbar\xi)\\
\hat{D}(z,s)\hat{f}(\hat{a},\hat{a}^{\dagger})\hat{D}^{-1}(z,s)&=&
\hat{f}(\hat{a}-z,\hat{a}^{\dagger}-\bar{z})
\end{eqnarray}
Making use of the relations given above and the
definitions (5) and (6) one can easily obtain
\begin{eqnarray}
\int \int\hat{\Delta}_{qp}(s)dqdp&=& h =h(2\pi a_{0}^{2})^{-1}
\int \int\hat{\Delta}_{Z}(s)d^{2}Z\\
Tr[\hat{\Delta}_{qp}(s)]&=& 1 =Tr[\hat{\Delta}_{Z}(s)]\\
\hat{\Delta}^{\dagger}_{qp}(s)&=& \hat{\Delta}_{qp}(-\bar{s})\qquad,\quad
\hat{\Delta}^{\dagger}_{Z}(s)=\hat{\Delta}_{Z}(\bar{s})
\end{eqnarray}
Now, two large class of associations can be defined as follows
\begin{eqnarray}
\hat{F}(\hat{q},\hat{p})&=&
h^{-1} \int\int f^{(-s)}(q,p)\hat{\Delta}_{qp}(s)dqdp\\
\hat{G}(\hat{a}^{\dagger},\hat{a})&=&
(2\pi a_{0}^{2})^{-1}\int\int g^{(-s)}(Z,\bar{Z})\hat{\Delta}_{Z}(s)d^{2}Z
\end{eqnarray}
whose inverse transformations are
\begin{eqnarray}
f^{(-s)}(q,p)&=&Tr[\hat{F}\hat{\Delta}_{qp}(-s)] \\
g^{(-s)}(Z,\bar{Z})&=&Tr[\hat{G}\hat{\Delta}_{Z}(-s)],
\end{eqnarray}  
respectively, where $Tr$ stands for the trace. 
For the special values $s=1,0,-1$, Eqs.(16) and
(18) are  known as
the standart, Weyl, and antistandard rules of associations, respectively. 
On the other hand, Eqs.(17) and (19) are known as the normal,  Weyl, and 
antinormal rules of associations for $s=1, 0, -1$ , respectively. 
If the density operator $\hat{\rho}$ of a quantum mechanical system
is mapped by Eqs.(18) and (19),  the resulting c-number functions
\begin{eqnarray}
W_{\rho}(q,p,-s)=Tr[\hat{\rho} \hat{\Delta}_{qp}(-s)] \qquad ,
\qquad W^{\prime}_{\rho}(Z,\bar{Z},-s)=Tr[\hat{\rho} \hat{\Delta}_{Z}(-s)]
\end{eqnarray}
are called, generically, the quasiprobability distribution 
functions (qdf). They enable us to carry out quantum
mechanical calculations in  classical manner 
in the corresponding phase space \cite{Lee,Hillery}. 
In the real parametrization, the qdf's corresponding to $s=0$ and $s=\mp1$
are called the Wigner and the Kirkwood qdf's, respectively. In  the
case of complex parametrization, the qdf corresponding to 
$s=1$ and $s=-1$ are known as the Glauber-Sudarshan
P-functions and Q-functions.
There is one more special association defined by
\begin{eqnarray}
\hat{F}(\hat{q},\hat{p})&=&
(\hbar /2\pi)\int \int f(\xi,\eta)\hat{D}^{-1}(\xi,\eta)d \xi d \eta \nonumber\\
f(\xi,\eta)&=&Tr[\hat{F} \hat{D}(\xi,\eta)].
\end{eqnarray}
This is known as the alternative Weyl association (or quantization), and
the above mentioned Wigner quantization is simply the Fourier transform
of it. The phase space resulting from Eq.(12) and having $(\xi,\eta)$
as  canonically conjugate coordinates, is also known as the 
Weyl phase space. In the case of  complex parametrization the
alternative Weyl association takes the form
\begin{eqnarray}
\hat{F}(\hat{a}^{\dagger},\hat{a})=
\pi^{-1}\int\int f(z,\bar{z})\hat{D}^{-1}(z,\bar{z})d^{2}z \qquad;\qquad
f(z,\bar{z})=Tr[\hat{F} \hat{D}(z,\bar{z})].
\end{eqnarray}

By using the properties of the $\hat{D}$ basis
it can be easily verified that the Hilbert-Schmid norm of an 
operator defined by $||\hat{F}||=(Tr[\hat{F}^{\dagger}\hat{F}])^{1/2}$
is equal to  the usual 
Hilbert space norm $||f||= (\langle f|f \rangle)^{1/2}$
of the corresponding c-number function. 
Thus, the alternative Weyl quantization is a norm preserving
$1-1$ association between the space of bounded operators and
the space of square integrable functions. Except for some particular
values of $s$ that  may give 
rise to singularities \cite{Cahill}, the other associations are
norm preserving $1-1$ associations as well.

The parametrized bases operators were for the first time, introduced by
Cahill and Glauber \cite{Cahill} in order to interpolate among various types
of orderings. Thus, the s-ordered
products $\hat{y}^{(s)}_{nm} \equiv
\{(\hat{a}^{\dagger})^{n}(\hat{a})^{m}\}_{s}$  and
$\hat{t}^{(s)}_{nm} \equiv \{(\hat{q})^{n}(\hat{p})^{m}\}_{s}$ are
defined as follows:
\begin{eqnarray}
\hat{y}^{(s)}_{nm}&=&
\partial^{n}_{z}\partial^{m}_{(-\bar{z})}\hat{D}(z,s)|_{z=0}\\
\hat{t}^{(s)}_{nm}&=&
(-i)^{n+m}\partial^{n}_{\xi}\partial^{m}_{\eta}\hat{D}(s)|_{\xi=0=\eta}
\end{eqnarray}
where, and henceforth, the notation
$\partial_{x}\equiv \partial / \partial x $.
will be used. By writing 
\begin{eqnarray}
\hat{D}(z,s)=e^{(s-s')|z|^{2}/2} \hat{D}(z,s') \qquad ;
\qquad \hat{D}(s)=e^{-i\hbar(s-s\prime)\xi \eta/2}\hat{D}(s\prime)
\end{eqnarray}
and differentiating, we obtain
\begin{eqnarray}
\hat{y}^{(s)}_{nm}&=&
\sum^{(n,m)}_{k=0}2^{-k}b(k,n,m)
[-(s-s')]^{k}\hat{y}^{(s')}_{n-k,m-k}\\
\hat{t}^{(s)}_{nm}&=&
\sum^{(n,m)}_{k=0}2^{-k}b(k,n,m)
[i\hbar(s-s')]^{k}\hat{t}^{(s')}_{n-k,m-k}
\end{eqnarray}
where $(n,m)$ denotes the smaller of the integers n and m, and
$(^{n}_{k})=n![(n-k)! k!]^{-1}$  being a binomial coefficient we set
\begin{eqnarray} 
b(k,n,m)=(^{n}_{k})(^{m}_{k})k!
\end{eqnarray}
These relations express an arbitrary s-ordered product in terms of a
polynomial in $s'$-ordered products  where $s'$ is 
also arbitrary. Note that the minus sign in $[-(s-s')]$ in Eq.(20)
and $i\hbar$ in Eq.(21). These  are the remnants of the
commutators of the corresponding operators there.
\section{DIFFERENTIAL STRUCTURE OF THE WEYL BASIS}
The embedding of orderings in a continuum provides a natural
context for viewing their differences and interrelationships
in a continuous manner and enable us to carry out the
related analyses in the most general form.
However, the definitions  (23) and (24) are quite implicit
in contrast with the simple notion of ordering as a
prescription about the arrangement of operators.
Moreover, some complicated and long formulas that  may be
encountered in the formulation of the WWGM-quantization can
be traced back to this implicit  definition of ordering.
The derivatives of
the bases $\hat{D}(s)$ and $\hat{D}(z,s)$
with respect to the phase space coordinates 
are obtained in this section.
An explicit formula for the ordered products which initiated the main
observations of this report will be derived
in the next section.

First, the following relations can be easily obtained from
(11) and (12):
\begin{eqnarray}
\xi \hat{D}(s)&=&\hbar^{-1}[\hat{p},\hat{D}(s)] \qquad ;
\qquad \eta \hat{D}(s)=-\hbar^{-1}[\hat{q},\hat{D}(s)]\\
z\hat{D}(z,s)&=&[\hat{a},\hat{D}(z,s)] \qquad ;
\qquad \bar{z} \hat{D}(z,s)=[\hat{a}^{\dagger},\hat{D}(z,s)]
\end{eqnarray}
They can be generalized as follows
\begin{eqnarray}
\xi^{n} \eta^{m}\hat{D}(s)=
(-{\hbar}^{-1} ad_{\hat{q}})^{m}({\hbar}^{-1} ad_{\hat{p}})^{n}\hat{D}(s) 
\qquad ; \qquad
z^{m}\bar{z}^{n}\hat{D}(z,s)=
(ad_{\hat{a}^{\dagger}})^{n}(ad_{\hat{a}})^{m}\hat{D}(z,s)
\end{eqnarray}
where $ad_{\hat{A}}$ denotes the adjoint action 
of $\hat{A}$: $ad_{\hat{A}}\hat{B}\equiv[\hat{A},\hat{B}]$.
Note that $[ad_{\hat{q}},ad_{\hat{p}}]=ad_{[\hat{q},\hat{p}]}=0$.
The same relation remains
valid if the pair $(\hat{q},\hat{p})$ is replaced by 
$(\hat{a}^{\dagger},\hat{a})$.
Thus, the ordering of ad operations in Eq.(31) is
inessential.

By taking the derivatives of the various factorizations
of $\hat{D}(s)$ and $\hat{D}(z,s)$
implied by the Baker-Campbell-Hausdorff (BCH)
formula we obtain the following identities:
\begin{eqnarray}
\partial_{\xi}\hat{D}(s)&=&i(\hat{q}+ \eta s^{-})\hat{D}(s)=
i\hat{D}(s)(\hat{q}-\eta s^{+})=
(i/2)[\hat{q}-\frac{1}{2}\hbar \eta s,\hat{D}(s)]_{+}\nonumber\\
\partial_{\eta}\hat{D}(s)&=&i(\hat{p}-\xi s^{+})\hat{D}(s)=
i\hat{D}(s)(\hat{p}+\xi s^{-})=
(i/2)[\hat{p}-\frac{1}{2}\hbar \xi s,\hat{D}(s)]_{+} \\
\partial_{z}\hat{D}(z,s)&=&[\hat{a}^{\dagger}-(\bar{z}s^{-}/ \hbar)]\hat{D}(z,s)=
\hat{D}(z,s)[\hat{a}^{\dagger}+(\bar{z}s^{+}/\hbar)]=
(1/2)[\hat{a}^{\dagger}+\frac{1}{2}\bar{z}s,\hat{D}(z,s)]_{+}\nonumber\\
\partial_{(-\bar{z})}\hat{D}(z,s)&=&[\hat{a}-(z s^{+}/\hbar)]\hat{D}(z,s)\nonumber\\
&=&\hat{D}(z,s)[\hat{a}+(z s^{-}/\hbar)]=(1/2)[\hat{a}-\frac{1}{2}zs,\hat{D}(z,s)]_{+}
\end{eqnarray}
where $[,]_{+}$ denotes the anticommutator and
\begin{eqnarray}
s^{\mp}=\frac{1}{2}\hbar (1 \mp s)
\end{eqnarray}
By making use of the Leibniz rule 
\begin{eqnarray}
\partial^{n}(u v)=\sum^{n}_{k=0}(^{n}_{k})(\partial^{k}u)(\partial^{n-k} v)
\end{eqnarray}
it is possible to obtain  generalizations of Eqs.(32)-(33) in several ways.
That is, depending on the way  the derivatives are taken,
the definitions (23) and (24) may not yield the desired s-ordered
product, but an equivalent one. In this sense
the definitions (23) and (24) are implicit.
The dual role played by the bases operators allows us to have
the opposite sides of Eqs.(29) and (30) contain quantities living
in different spaces. On the other hand Eqs.(32)-(33) are not written 
in the same way.
The implicit nature of the above formulas is due this fact.
Eqs.(32)-(33) can  be rewritten in such a way that the quantities
appearing on the opposite sides of the equalities live in different 
spaces. This we may achieve in two ways: (i) by taking the terms 
proportional to $\eta \hat{D}(s)$, $\xi \hat{D}(s)$, $z \hat{D}(z,s)$, 
and $\bar{z}\hat{D}(z,s)$ to the left, hence
leaving all the Hilbert space quantities at the right, or (ii)
by replacing $\hbar \eta \hat{D}(s)$, $\hbar \xi \hat{D}(s)$, 
$z\hat{D}(z,s)$, and $\bar{z} \hat{D}(z,s)$  in view of Eqs.(29)-(30)
by $-ad_{\hat{q}}$, $ad_{\hat{p}}$, $ad_{\hat{a}}$, and 
$ad_{\hat{a}^{\dagger}}$ , respectively. The first way leads to the
introduction of s-parametrized Bopp operators that we are 
going to investigate at the end of this section. The second way allows us to rewrite
Eqs.(32)-(33) in the following unique form:
\begin{eqnarray}
\partial_{\xi}\hat{D}(s)&=&(i/2)\hat{T}_{[\hat{q}]_{(s)}}\hat{D}(s) \qquad ;
\qquad \partial_{\eta}\hat{D}(s)=(i/2)\hat{T}_{[\hat{p}]_{(-s)}}\hat{D}(s) \\
\partial_{z}\hat{D}(z,s)&=&(1/2)\hat{T}_{[\hat{a}^{\dagger}]_{(s)}}\hat{D}(z,s)  ;
 \partial_{(-\bar{z})}\hat{D}(z,s)=
(1/2)\hat{T}_{[\hat{a}]_{(-s)}}\hat{D}(z,s)
\end{eqnarray}
where  we define the Hilbert space operation
\begin{eqnarray}
\hat{T}_{[\hat{A}]_{(s)}}=(1+s)\hat{L}_{\hat{A}}+(1-s)\hat{R}_{\hat{A}}.
\end{eqnarray}
Here $\hat{L}_{\hat{A}}$ and $\hat{R}_{\hat{A}}$
are, respectively, the multiplication from left and from 
right by $\hat{A}$. In fact $\hat{T}_{[\hat{A}]_{(-s)}}$
 is an s-deformation of 
$\hat{T}_{[\hat{A}]_{+}}\equiv \hat{L}_{\hat{A}}+\hat{R}_{\hat{A}} $.
It is equal to $2\hat{L}_{\hat{A}}$, $\hat{T}_{[\hat{A}]_{+}}$ and
$2\hat{R}_{\hat{A}}$ for $s=1$, $s=0$, and $s=-1$, respectively.
We observe that for an arbitrary operator $\hat{B}$
\begin{eqnarray}
[\hat{T}_{[\hat{q}]_{(s)}} ,\hat{T}_{[\hat{p}]_{(-s)}}]\hat{B} =
0 = [\hat{T}_{[\hat{a}^{\dagger}]_{(s)}} , \hat{T}_{[\hat{a}]_{(-s)}}]\hat{B}
\end{eqnarray}
which simply follows from 
$\partial_{\xi}\partial_{\eta}=\partial_{\eta}\partial_{\xi}$ and
the relations (36), (37). It can also be directly verified.
Then we  generalize Eqs.(36),(37) as follows:
\begin{eqnarray}
\partial^{n}_{\xi} \partial^{m}_{\eta}\hat{D}(s)&=&
(i/2)^{n+m}\hat{T}^{n}_{[\hat{q}]_{(s)}}
\hat{T}^{m}_{[\hat{p}]_{(-s)}}\hat{D}(s) \\
\partial^{n}_{z}\partial^{m}_{(-\bar{z})}\hat{D}(z,s)&=&
2^{-(n+m)}\hat{T}^{n}_{[\hat{a}^{\dagger}]_{(s)}}
\hat{T}^{m}_{[\hat{a}]_{(-s)}}\hat{D}(z,s)
\end{eqnarray}
In view of Eqs.(39)-(41), these can be rewritten in a finitely many
different looking but equivalent forms.

We substitute Eqs.(40) and (41) in the definitions (23) and (24) to obtain
\begin{eqnarray}
\hat{t}^{(s)}_{nm}&=&
2^{-(n+m)}\hat{T}^{n}_{[\hat{q}]_{(s)}}
\hat{T}^{m}_{[\hat{p}]_{(-s)}}\hat{I}=
2^{-(n+m)}\hat{T}^{m}_{[\hat{p}]_{(-s)}}
\hat{T}^{n}_{[\hat{q}]_{(s)}}\hat{I},\\
\hat{y}^{(s)}_{nm}&=&
2^{-(n+m)}\hat{T}^{n}_{[\hat{a}^{\dagger}]_{(s)}}
\hat{T}^{m}_{[\hat{a}]_{(-s)}}\hat{I}=
2^{-(n+m)}\hat{T}^{m}_{[\hat{a}]_{(-s)}}
\hat{T}^{n}_{[\hat{a}^{\dagger}]_{(s)}}\hat{I}.
\end{eqnarray}
By making use of the binomial formula
\begin{eqnarray}
\hat{T}^{n}_{[\hat{A}]_{(s)}} \equiv
[(1+s)\hat{L}_{\hat{A}}+(1-s)\hat{R}_{\hat{A}}]^{n}=
\sum^{n}_{j=0}(^{n}_{j}) (1+s)^{j}(1-s)^{n-j}
\hat{L}^{j}_{\hat{A}}\hat{R}^{n-k}_{\hat{A}}
\end{eqnarray} 
we can rewrite  expressions in (44) more explicitly as 
\begin{eqnarray}
\hat{t}^{(s)}_{nm}&=&2^{-n}\sum^{n}_{j=0}(^{n}_{j})(1+s)^{j}(1-s)^{n-j}
\hat{q}^{j}\hat{p}^{m}\hat{q}^{n-j}\\
&=&2^{-m}\sum^{m}_{k=0}(^{m}_{k})(1-s)^{k}(1+s)^{m-k}
\hat{p}^{k}\hat{q}^{n}\hat{p}^{m-k}.
\end{eqnarray}
A similar relation holds for $\hat{y}^{(s)}_{nm}$ if 
the pair $(\hat{q},\hat{p})$ 
is replaced by $(\hat{a}^{\dagger},\hat{a})$ in these relations. 
In view of Eq.(39), it is possible to write many equivalent forms of the 
above relations. But for later use we have written only two of them. 
From these we get for $s=\pm1$
\begin{eqnarray}
\hat{t}^{(1)}_{nm}=\hat{L}^{n}_{\hat{q}}\hat{R}^{m}_{\hat{p}}\hat{I}=
\hat{q}^{n}\hat{p}^{m}\qquad;\qquad \hat{t}^{(-1)}_{nm}=
\hat{L}^{m}_{\hat{p}}\hat{R}^{n}_{\hat{q}}\hat{I}=\hat{p}^{m}\hat{q}^{n}
\end{eqnarray}
and for $s=0$
\begin{eqnarray}
\hat{t}^{(0)}_{nm}&=&2^{-n}\sum^{n}_{j=0}(^{n}_{j})\hat{q}^{j}\hat{p}^{m}
\hat{q}^{n-j}\\
&=&2^{-m}\sum^{m}_{k=0}(^{m}_{k})\hat{p}^{k}\hat{q}^{n}\hat{p}^{m-k}
\end{eqnarray}
The expressions  (47) exhibit the standart and antistandart
rules of ordering while  (48) and (49) yield the two well
known expressions for symmetrically (or Weyl) ordered products. 
In fact the usual expression known for the Weyl ordered form of
$\hat{t}^{(0)}_{nm}$ is a totally symmetrized product containing n factors 
of $\hat{q}$ and m factors of $\hat{p}$, 
normalized by dividing by the number of terms in the 
symmetrized expression. 
In the literature \cite{Bender}
the equivalence of these
three Weyl ordered forms is said to be verified by using
the usual commutation relations.  This requires
long and tedious computations. In our formulation on the other hand,
not  only the above mentioned equivalences, but also   explicit expressions 
for many forms of the Weyl ordered products arise
naturally as a corollary to Eq.(39).

As an application, let us consider the traces of Eqs.(40)
and (41) for $s=0$. By
noting that $Tr\hat{D}=(2\pi / \hbar) \delta(\xi)\delta(\eta)$,
$Tr\hat{D}(z)=\pi \delta^{2}(z)$ 
where $\hat{D}\equiv\hat{D}(0)$;  
the well known  Weyl associations follow:
\begin{eqnarray}
\frac{2 \pi}{\hbar}\partial^{n}_{\xi}\delta(\xi)
\partial^{m}_{\eta}\delta(\eta)
&\leftrightarrow&(i)^{n+m}\hat{t}^{(0)}_{nm}\\
\pi \partial^{n}_{z}\partial^{m}_{(-\bar{z})}\delta^{2}(z)
&\leftrightarrow&\hat{y}^{(0)}_{nm}
\end{eqnarray}
These expressions cannot be so easily obtained in other approaches. 

We propose that the relation (45) (or alternatively (46)) can
be given as the definition
of  s-ordered product. There are several reasons that support
this suggestion:
(i) These expressions are simpler and more explicit
than the definitions given by (23) and (24).
Neither the phase space coordinates nor
the basis operators  appear in these expressions. 
(ii) The ordered products
of $\hat{a}^{\dagger}$ and $\hat{a}$ can be treated
on an equal footing. In fact the disapperance of  
$\hbar$, or any multiple of it, in these expressions implies that
all the relations and the remarks given above
are valid for any algebra having
$[\hat{A,\hat{B}}]=i\lambda \hat{I};\lambda \in \bf C$ \cite{Gelfand}.
A physical application may be the algebra of 
velocity operators of a charged particle 
moving in an external electromagnetic field.
(iii) These definitions may be extended to
 the case when one or both of the integers $n$ and $m$
are negative.
Furthermore, by using these relations the Hermiticity
of a general s-ordered product can be easily decided.
From (45) and (46) it follows that
$[\hat{t}^{(s)}_{nm}]^{\dagger} = \hat{t}_{nm}^{(-\bar{s})}$.
Thus, for every pair of integers n,m , $\hat{t}^{(s)}_{nm}$
are Hermitian provided $\bar{s}=-s$. In particular,
the Weyl ordered products $\hat{t}^{(0)}_{nm}$ are Hermitian. 
In the case of
$\hat{a}, \hat{a}^{\dagger}$ we have, from (45) and (46),
$[\hat{y}^{(s)}_{nm}]^{\dagger}=\hat{y}_{mn}^{(\bar{s})}$. Hence,
the s-ordered products $\hat{y}^{(s)}_{nm}$ are Hermitian
if and only if $m=n$ , and $\bar{s}=s$. For general
$s \in \bf {C}$ one can find combinations such as
\begin{eqnarray}
\hat{\kappa}_{nm}(s) = \alpha \hat{t}_{nm}^{(s)} +
\bar{\alpha}\hat{t}_{nm}^{(-\bar{s})} \qquad,\qquad
\hat{\kappa}^{\prime}_{nm}(s) =
\alpha \hat{y}_{nm}^{(s)} + \bar{\alpha}\hat{y}_{mn}^{(\bar{s})}
\end{eqnarray}
$(\alpha \in \bf {C})$ that are Hermitian.


We now consider the alternative way of writing the derivatives
in $\hat{D}(s)$  so that the quantities appearing at opposite
sides  belong to different spaces as
\begin{eqnarray}
(-i\partial_{\xi}-s^{-}\eta)\hat{D}(s)&=&
\hat{q}\hat{D}(s)\qquad,\qquad (-i\partial_{\xi}+s^{+}\eta)\hat{D}(s) =
\hat{D}(s)\hat{q}\nonumber\\ 
(-i\partial_{\eta}+s^{+}\xi)\hat{D}(s)&=&\hat{p}\hat{D}(s)\qquad,
\qquad (-i\partial_{\eta}-s^{-}\xi)\hat{D}(s) = \hat{D}(s)\hat{p} 
\end{eqnarray}                                            
By defining the s-parametrized Bopp operators
\begin{eqnarray}  
Q_{L}(s)=-i\partial_{\xi} - s^{-}\eta \qquad, \qquad Q_{R}(s)=
-i\partial_{\xi} + s^{+}\eta \nonumber\\
P_{L}(s)=-i\partial_{\eta} + s^{+}\xi \qquad, \qquad P_{R}(s)=
-i\partial_{\eta} - s^{-}\xi
\end{eqnarray}
we can generalize Eqs.(46):
\begin{eqnarray}  
Q_{L}^{n}(s)\hat{D}(s)&=&\hat{q}^{n}\hat{D}(s)\qquad,
\qquad Q_{R}^{n}(s)\hat{D}(s) = \hat{D}(s)\hat{q}^{n}\nonumber\\ 
P_{L}^{n}(s)\hat{D}(s)&=&
\hat{p}^{n}\hat{D}(s)\qquad,\qquad P_{R}^{n}(s)\hat{D}(s) =
\hat{D}(s)\hat{p}^{n} 
\end{eqnarray}
Being defined in the tangent space of the $(\xi,\eta)$-phase
space, the s-parametrized Bopp operators obey the commutation 
relations
\begin{eqnarray}  
[Q_{L}(s),P_{L}(s)]=-i\hbar =-[Q_{R}(s),P_{R}(s)] 
\end{eqnarray}
with all the other commutators equal to  zero. These relations indicate that 
the s-parametrized Bopp operators provide a concrete coordinate
realization of the direct sum of two copies of the HW-algebra, and
for real $s$ they are hermitian on the Lebesque space defined on the
$(\xi,\eta)$-phase space. 

In the case of complex coordinates  Eqs.(37) yield
\begin{eqnarray}  
Q^{\prime n}_{L}(s)\hat{D}(z,s)&=&
(\hat{a}^{\dagger})^{n}\hat{D}(z,s) \qquad,\qquad
Q^{\prime n}_{R}(s)\hat{D}(z,s) =
\hat{D}(s)(\hat{a}^{\dagger})^{n}\nonumber\\
P^{\prime n}_{L}(s)\hat{D}(z,s)&=&(-\hat{a})^{n}\hat{D}(z,s) \qquad,
 P^{\prime n}_{R}(s)\hat{D}(z,s) = \hat{D}(z,s)(-\hat{a})^{n} 
\end{eqnarray}
where we have defined the s-parametrized complex Bopp operators
\begin{eqnarray}  
Q^{\prime}_{L}(s)&=&\partial_{z} + \bar{z}(s^{-}/\hbar)\qquad,\qquad
Q^{\prime}_{R}(s) = \partial_{z}-\bar{z}(s^{+}/ \hbar)\nonumber\\
P^{\prime}_{L}(s)&=&\partial_{\bar{z}}- z(s^{+}/ \hbar)\qquad,\qquad
P^{\prime}_{R}(s) =  \partial_{\bar{z}}+ z(s^{-}/\hbar).
\end{eqnarray}
The nonvanishing commutation relations they satisfy  are
\begin{eqnarray}  
[Q^{\prime}_{L}(s),P^{\prime}_{L}(s)]=-I=
-[Q^{\prime}_{R}(s),P^{\prime}_{R}(s)] 
\end{eqnarray}
We note that the complex Bopp operators are related 
with a coordinate realization 
of the bosonic annihilation and creation operators.
We wish  also to remark in passing that 
Eqs.(55) and (57) resemble eigenvalue equations.

The Bopp operators were originally defined only for the Wigner (s=0)
quantization \cite{Hillery,Bopp}. 
They play an important role in the derivative-based approach that we are using.
So  we generalize them  for any quantization rule.
\section{DIFFERENTIAL STRUCTURE OF THE WIGNER BASIS}
Differential structure of the $\hat{\Delta}(s)$ bases are formally 
the Fourier transform of that obtained in the preceding 
sections  for the $\hat{D}(s)$ bases. To put it more  
simply, they can be derived  from the definition (5), (6)
by elementary calculations. Indeed, making use of Eqs.(29) and (30), it is
easy to verify that  
\begin{eqnarray}
\partial_{q} \hat{\Delta}_{qp}(s)&=&
-\frac{i}{\hbar} [\hat{p}, \hat{\Delta}_{qp}(s)]
\qquad,\qquad \partial_{p} \hat{\Delta}_{qp}(s) =
\frac{i}{\hbar} [\hat{q}, \hat{\Delta}_{qp}(s)] \nonumber\\
\partial_{Z} \hat{\Delta}_{Z}(s)& =&
\frac{1}{a_{0}\sqrt{2}} [\hat{a}^{\dagger},
 \hat{\Delta}_{Z}(s)] \qquad,\qquad
 \partial_{\bar{Z}} \hat{\Delta}_{Z}(s) =
 -\frac{1}{a_{0}\sqrt{2}} [\hat{a}, \hat{\Delta}_{Z}(s)]
\end{eqnarray}
and from (36) and (38) it follows that 
\begin{eqnarray}
q\hat{\Delta}_{qp}(s)&=&
\frac{1}{2}\hat{T}_{[\hat{q}]_{(s)}}\hat{\Delta}_{qp}(s)
\qquad,\qquad p\hat{\Delta}_{qp}(s) =
\frac{1}{2} \hat{T}_{[\hat{p}]_{(-s)}}\hat{\Delta}_{qp}(s) \nonumber\\
Z\hat{\Delta}_{Z}(s)&=&
\frac{a_{0}}{\sqrt{2}}\hat{T}_{[\hat{a}]_{(-s)}}\hat{\Delta}_{Z}(s)
\qquad, \qquad \bar{Z} \hat{\Delta}_{Z}(s)=
\frac{a_{0}}{\sqrt{2}}\hat{T}_{[\hat{a}^{\dagger}]_{(s)}}\hat{\Delta}_{Z}(s).
\end{eqnarray}
This is the only place where we need partial integration.
Recalling the commutation relations (39), the above expressions may 
be generalized as follows:
\begin{eqnarray}
q^{n}p^{m}\hat{\Delta}_{qp}(s)&=&
2^{-(n+m)}\hat{T}^{n}_{[\hat{q}]_{(s)}}
\hat{T}^{m}_{[\hat{p}]_{(-s)}}\hat{\Delta}_{qp}(s) \nonumber\\
\bar{Z}^{n}Z^{m}\hat{\Delta}_{Z}(s)&=&
(\frac{a_{0}}{\sqrt{2}})^{n+m}
\hat{T}^{n}_{[\hat{a}^{\dagger}]_{(s)}}
\hat{T}^{m}_{[\hat{a}]_{(-s)}}\hat{\Delta}_{Z}(s).
\end{eqnarray}

As an example, by taking the traces of both sides of  
Eqs.(62) we obtain
\begin{eqnarray}
q^{n}p^{m}&=&Tr[\hat{t}_{nm}^{(s)}\hat{\Delta}_{qp}(-s)]\nonumber\\
\bar{Z}^{n}Z^{m}&=&
(a_{0}\sqrt {2})^{n+m}
Tr[\hat{y}^{(s)}_{nm}\hat{\Delta}_{Z}(-s)].
\end{eqnarray}
Alternatively, taking the integrals of both sides of
the same equations we are led to 
\begin{eqnarray}
\hat{t}^{(s)}_{nm} &=&
h^{-1}\int \int q^{n}p^{m}\hat{\Delta}_{qp}(s) dqdp , \nonumber\\
\hat{y}^{(s)}_{nm}&=&
(2\pi a^{2}_{0})^{-1}(a_{0}\sqrt{2})^{-(n+m)}
\int \int \bar{Z}^{n}Z^{m}\hat{\Delta}_{Z}(s) d^{2}Z.
\end{eqnarray}
Eqs.(63) and (64) explicitly show that the s-quantization of the monomials
$q^{n}p^{m}$ and $\bar{Z}^{n}Z^{m}(a_{0}\sqrt{2})^{-(n+m)}$ are
nothing but the s-ordered products
$\hat{t}_{nm}^{(s)}$ and $\hat{y}_{nm}^{(s)}$, respectively.
In our approach these well known results concerning 
the WWGM-quantization are obtained with ease
in a unified manner. 

From Eqs.(60) and (61) we have
\begin{eqnarray}
Q_{\Delta L}(s)\hat{\Delta}_{qp}(s)&=&\hat{q}\hat{\Delta}_{qp}(s) \qquad,
\qquad Q_{\Delta R}(s)\hat{\Delta}_{qp}(s)=\hat{\Delta}_{qp}(s)\hat{q}\nonumber\\
P_{\Delta L}(s)\hat{\Delta}_{qp}(s)&=&\hat{p}\hat{\Delta}_{qp}(s) \qquad,
\qquad P_{\Delta R}(s)\hat{\Delta}_{qp}(s)=\hat{\Delta}_{qp}(s)\hat{p}
\end{eqnarray}
where
\begin{eqnarray}
Q_{\Delta L}(s)&=&q - is^{-}\partial_{p}\qquad,Q_{\Delta R}(s)=
q + is^{+}\partial_{p} \nonumber\\
P_{\Delta L}(s)&=&p + is^{+}\partial_{q} \qquad , \qquad P_{\Delta R}(s)=
p - is^{-}\partial_{q}
\end{eqnarray}
are the s-parametrized Bopp operators for the $\hat{\Delta}_{qp}(s)$ bases.
The only nonvanishing commutators for them are
\begin{eqnarray}
[Q_{\Delta L}(s),P_{\Delta L}(s)]=-i\hbar=-[Q_{\Delta R}(s),P_{\Delta R}(s)].
\end{eqnarray}

We also give  the s-parametrized complex Bopp operators
\begin{eqnarray}
Q^{\prime}_{\Delta L}(s)=
2^{-1/2}[\frac{\bar{Z}}{a_{0}}+a_{0}(1-s)\partial_{Z}]\qquad ,\qquad
Q^{\prime}_{\Delta R}(s)=
2^{-1/2}[\frac{\bar{Z}}{a_{0}}-a_{0}(1+s)\partial_{Z}]\nonumber\\ 
P^{\prime}_{\Delta L}(s)=
2^{-1/2}[\frac{Z}{a_{0}}-a_{0}(1+s)\partial_{\bar{Z}}]\qquad ,\qquad
P^{\prime}_{\Delta R}(s)=
2^{-1/2}[\frac{Z}{a_{0}}+a_{0}(1-s)\partial_{\bar{Z}}] 
\end{eqnarray}
Their action on bases  can be obtained from (60) and (61) as
\begin{eqnarray}
Q^{\prime n}_{\Delta L}(s)\hat{\Delta}_{Z}(s)&=&
(\hat{a}^{\dagger})^{n}\hat{\Delta}_{Z}(s)\qquad,\qquad
Q^{\prime n}_{\Delta R}(s)\hat{\Delta}_{Z}(s)=
\hat{\Delta}_{Z}(s)(\hat{a}^{\dagger})^{n}\nonumber\\
P^{\prime n}_{\Delta L}(s)\hat{\Delta}_{Z}(s)&=&
(\hat{a})^{n}\hat{\Delta}_{Z}(s)\qquad,\qquad
P^{\prime n}_{\Delta R}(s)\hat{\Delta}_{Z}(s)=
\hat{\Delta}_{Z}(s)(\hat{a})^{n}.
\end{eqnarray}
In the complex case the nonvanishing commutators are
\begin{eqnarray}
[Q^{\prime }_{\Delta L}(s),P^{\prime }_{\Delta L}(s)]=1=
-[Q^{\prime}_{\Delta R}(s),P^{\prime }_{\Delta R}(s)].
\end{eqnarray}

We will finish this section by another important observation 
that generalizes  a well known relation between the
Wigner $(s=0)$ association  and arbitrary s-association.
It follows  immediately from  
\begin{eqnarray}
\hat{t}_{nm}^{(r)}\hat{\Delta}_{qp}(s)&=&
\{Q^{n}_{\Delta L}(s)P^{m}_{\Delta L}(s)\}_{-r}\hat{\Delta}_{qp}(s)\nonumber\\
\hat{\Delta}_{qp}(s)\hat{t}_{nm}^{(r)}&=&
\{Q^{n}_{\Delta R}(s)P^{m}_{\Delta R}(s)\}_{r}\hat{\Delta}_{qp}(s)
\end{eqnarray}
by taking trace of both sides and making use of Eqns. (14), (45), (46) and (65),
\begin{eqnarray}
Tr[\hat{t}_{nm}^{(r)}\hat{\Delta}_{qp}(s)]=
\{Q^{n}_{\Delta L}(s)P^{m}_{\Delta L}(s)\}_{-r}I=
\{Q^{n}_{\Delta R}(s)P^{m}_{\Delta R}(s)\}_{r}I.
\end{eqnarray}
Thus  for any arbitrary $r$ and $s$,
the c-number function corresponding to an r-ordered 
product via the s-rule of association can be obtained by the
action of the r-ordered s-parametrized Bopp
operators on the phase space identity operator $I$.
The extension of these observations to the case of
complex coordinates and by linearity to arbitrary functions
of operators that can be expanded to a series of  ordered products 
is straightforward.
\section{$W_{\infty}$-COVARIANCE OF THE BASES OPERATORS}
By following the same lines leading to Eqs.(71) from
(45) (or (46)) and (55) we obtain
\begin{eqnarray}
T^{(r)}_{nm}(s)\hat{D}(s) = [\hat{t}_{nm}^{(r)},\hat{D}(s)]
\end{eqnarray}
where 
\begin{eqnarray}
T^{(r)}_{nm}(s) \equiv
\{Q^{n}_{L}(s)P^{m}_{L}(s)\}_{-r}-\{Q^{n}_{R}(s)P^{m}_{R}(s)\}_{r}.
\end{eqnarray}
Equation (73) reveals the important fact that each r-ordered product
of  $\hat{q}$'s and  $\hat{p}$'s generates an infinitesimal 
transformation in the Weyl basis $\hat{D}(s)$ in the 
Hilbert space $\cal{H}$. This transformation
 corresponds to an infinitesimal transformation 
of the basis in the Weyl   
phase space that is built up in terms of
$r$ and $-r$ ordered Bopp operators. 
The exponentiation of these transformations leads to
\begin{eqnarray}
V_{nm}(r,s)\hat{D}(s)=
\hat{U}_{nm}(r)\hat{D}(s)\hat{U}^{-1}_{nm}(r)
\end{eqnarray}
where $\gamma_{nm} \in \bf {C}$ are the transformation parameters and
\begin{eqnarray}
V_{nm}(r,s)
\equiv \exp(i\gamma_{nm}T^{(r)}_{nm}(s))\qquad,\qquad
\hat{U}_{nm}(r) \equiv \exp(i\gamma_{nm}\hat{t}_{nm}^{(r)}).
\end{eqnarray}
$\hat{U}^{-1}_{nm}(r)$ denotes the operator inverse of
$\hat{U}_{nm}(r)$.
Eq.(73) at the algebra level and  Eq.(75)
at the group level explain what we mean in its full generality
by $W_{\infty}$-covariance of the Weyl basis. Here we have 
two $W_{\infty}$ algebras (and groups): First one is generated
by the ordered products $\hat{t}^{(r)}_{nm}; n,m \geq 0$
and is acting in $\cal {H}$. The second is generated by
$T^{(r)}_{nm}(s); n,m \geq0 $ and is  realized in the  
Weyl phase space. 

Let $s=0$ and multiply both sides of Eqs.(73) and (75) by
an arbitrary bounded operator 
$\hat{F}(\hat{q},\hat{p})$. 
Then take the trace of the resulting equations. We thus obtain
\begin{eqnarray}
T^{(r)}_{nm}f(\xi,\eta)&=&Tr([\hat{F}(\hat{q},\hat{p}),\hat{t}_{nm}^{(r)}]
\hat{D})\\
V_{nm}(r,0)f(\xi,\eta)&=&Tr\{[\hat{U}^{-1}_{nm}(r)
\hat{F}(\hat{q},\hat{p})\hat{U}_{nm}(r)]\hat{D}\}
\end{eqnarray}
where $f=Tr[\hat{F}\hat{D}]$, $T^{(r)}_{nm} \equiv T^{(r)}_{nm}(0)$.
These two equations describe the $W_{\infty}$-covariance of the
Weyl quantization both at the algebra level (Eq.(77)) and at the group
level (Eq.(78)). In other words, if $\hat{F}$ is the Weyl quantization of
a c-number function $f$, then the $W_{\infty}$ transform of $\hat{F}$,
\begin{eqnarray}
\hat{F}^{\prime}=\hat{U}^{-1}_{nm}(r)\hat{F}\hat{U}_{nm}(r)
\end{eqnarray}
is the Weyl quantization of the c-number function
\begin{eqnarray}
f^{\prime}(\xi,\eta)=V_{nm}(r,0)f(\xi,\eta).
\end{eqnarray}

For the sake of brevity, in the case of complex coordinates
we write out only the main equations:
\begin{eqnarray}
T^{\prime (r)}_{nm}(s)\hat{D}(z,s)&=&
[\hat{y}^{(r)}_{nm},\hat{D}(z,s)]\nonumber\\
V^{\prime}_{nm}(r,s)\hat{D}(z,s)&=&
\hat{U}^{\prime}_{nm}(r)\hat{D}(z,s)\hat{U}^{\prime -1}_{nm}(r)
\end{eqnarray}
and
\begin{eqnarray}
T^{\prime (r)}_{nm}f(z,\bar{z})&=&
Tr\{[\hat{F}(\hat{a}^{\dagger},\hat{a}),\hat{y}^{(r)}_{nm}]\hat{D}(z)\}\nonumber\\
V_{nm}^{\prime}(r,0)f(z,\bar{z})&=&
Tr\{[\hat{U}^{\prime -1}_{nm}(r)
\hat{F}(\hat{a}^{\dagger},\hat{a})\hat{U}^{\prime }_{nm}(r)]\hat{D}(z)\}.
\end{eqnarray}
These expressions describe the $W_{\infty}$-covariance of 
the $\hat{D}(z,s)$ basis and of the Weyl quantization, respectively. 
Here we used the abrreviations $(\alpha_{nm}\in {\bf C})$
\begin{eqnarray}
T^{\prime (r)}_{nm}(s) & \equiv &
\{Q^{\prime n}_{L}(s)P^{\prime m}_{L}(s)\}_{-r}-
\{Q^{\prime n}_{R}(s)P^{\prime m}_{R}(s)\}_{r}\nonumber\\
V^{\prime}_{nm}(r,s) & \equiv &
\exp(i\alpha_{nm}T^{\prime (r)}_{nm}(s))\nonumber\\
\hat{U}^{\prime}_{nm}(r) & \equiv & \exp(i\alpha_{nm}\hat{y}_{nm}^{(r)})
\end{eqnarray}
and $ T^{\prime (r)}_{nm} \equiv T^{\prime (r)}_{nm}(0)$.

From Eq.(71) we have
\begin{eqnarray}
\Gamma^{(r)}_{nm}(s) \hat{\Delta}_{qp}(s)=
[\hat{t}^{(r)}_{nm}, \hat{\Delta}_{qp}(s)]
\end{eqnarray}
where
\begin{eqnarray}
\Gamma^{(r)}_{nm}(s)=
\{Q^{n}_{\Delta L}(s)P^{m}_{\Delta L}(s)\}_{-r} -
\{Q^{n}_{\Delta R}(s)P^{m}_{\Delta R}(s)\}_{r}.
\end{eqnarray}
It is straightforward to verify that in the case of  complex 
coordinates the corresponding relations are the following:
\begin{eqnarray}
\Gamma^{\prime (r)}_{nm}(s)\hat{\Delta}_{Z}(s)&=&
[\hat{y}^{(r)}_{nm},\hat{\Delta}_{Z}(s)]\\
\Gamma^{\prime (r)}_{nm}(s)&=&
\{Q^{\prime n}_{\Delta L}(s)P^{\prime m}_{\Delta L}(s)\}_{-r} -
\{Q^{\prime n}_{\Delta R}(s)P^{\prime m}_{\Delta R}(s)\}_{r}\nonumber
\end{eqnarray}
Exponentiating the actions (84) and (86) we are led to
\begin{eqnarray}
V^{\Delta}_{nm}(r,s)\hat{\Delta}_{qp}(s)&=&
\hat{U}_{nm}(r)\hat{\Delta}_{qp}(s)\hat{U}^{-1}_{nm}(r)\nonumber\\
V^{\prime \Delta}_{nm}(r,s)\hat{\Delta}_{Z}(s)&=&
\hat{U}^{\prime}_{nm}(r)
\hat{\Delta}_{Z}\hat{U}^{\prime -1}_{nm}(r)
\end{eqnarray}
where
\begin{eqnarray}
V^{\Delta}_{nm}(r,s)
\equiv\exp(i\gamma_{nm}\Gamma^{(r)}_{nm}(s))\qquad,\qquad
V^{\prime \Delta}_{nm}(r,s)
\equiv\exp(i\alpha_{nm}\Gamma^{\prime (r)}_{nm})\nonumber
\end{eqnarray}
These expressions exhibit both at the algebra and at the group level,
the $W_{\infty}$-covariance of the Wigner ($s=0$) and the Kirkwood
($s=\pm 1$) bases. Suppose $\hat{F}(\hat{q},\hat{p})$
and $\hat{G}(\hat{a}^{\dagger},\hat{a})$
are two arbitrary bounded operators, and let
$\hat{F}^{\prime}=\hat{U}^{-1}_{nm}(r)\hat{F} \hat{U}_{nm}(r) $ and 
$\hat{G}^{\prime}=
\hat{U}^{\prime -1}_{nm}(r)\hat{G} \hat{U}^{\prime}_{nm}(r) $
be their $W_{\infty}$ transforms. Then from Eqs.(87) we get
\begin{eqnarray}
V^{\Delta}_{nm}(r,s)f^{(s)}(q,p)=
Tr[\hat{F}^{\prime}\hat{\Delta}_{qp}(s)] \nonumber\\
V^{\prime \Delta}_{nm}(r,s)g^{(s)}(Z,\bar{Z})=
Tr[\hat{G}^{\prime}\hat{\Delta}_{Z}(s)]
\end{eqnarray}
where $f^{(s)}(q,p)=Tr[\hat{F}\hat{\Delta}_{qp}(s)]$ and 
$g^{(s)}(Z,\bar{Z})=Tr[\hat{G}\hat{\Delta}_{Z}(s)]$
are the corresponding c-number functions. The infinitesimal version
of (84) are
\begin{eqnarray}
\Gamma^{(r)}_{nm}(s)f^{(s)}(q,p)&=&
Tr([\hat{F},\hat{t}^{(r)}_{nm}]\hat{\Delta}_{qp}(s)) \nonumber\\
\Gamma^{\prime (r)}_{nm}(s)g^{(s)}(Z,\bar{Z})&=&
Tr([\hat{G},\hat{y}^{(r)}_{nm}]\hat{\Delta}_{Z}(s)).
\end{eqnarray}
Thus, the complete
$W_{\infty}$-covariance of the WWGM-quantization
is achieved. Explicit expressions giving the algebra generators
for $n,m \leq 2$ are presented below.
\begin{eqnarray}
\hat{t}_{00}^{(0)} &=& \hat{I} \qquad,\qquad \Gamma^{(0)}_{00}(s)= 0  \nonumber \\
 \hat{t}_{10}^{(0)}&=& \hat{q} \qquad,\qquad \Gamma^{(0)}_{10}(s) = -i\hbar\partial_{p} \nonumber \\
\hat{t}_{01}^{(0)}&=& \hat{p} \qquad,\qquad \Gamma^{(0)}_{01}(s)= i\hbar\partial_{q} \nonumber \\
\hat{t}_{11}^{(0)}&=& \frac{1}{2}(\hat{q}\hat{p}+\hat{p}\hat{q}) \qquad,\qquad\Gamma^{(0)}_{11}(s)=
i\hbar(q\partial_{q}-p\partial_{p}) \nonumber \\
\hat{t}_{20}^{(0)}&=& \hat{q}^{2}\qquad,\qquad \Gamma^{(0)}_{20}(s)= 
-2i\hbar q\partial_{p}+s\hbar^{2}\partial^{2}_{p} \nonumber \\
\hat{t}_{02}^{(0)}&=& \hat{p}^{2}\qquad, \qquad \Gamma^{(0)}_{02}(s)=
2i\hbar p\partial_{q}-s\hbar^{2}\partial^{2}_{q}  
\end{eqnarray}


In order to see the connection between the algebra of 
canonical diffeomorphisms and the $W_{\infty}$-algebra found
above in a different bases, we consider  the relation \cite{vercin}
\begin{eqnarray}
\Gamma^{(s)}_{nm}(-s)f(q,p)&=&
(q^{n}p^{m})\star_{(-s)}f(q,p)-f(q,p)\star_{(-s)}(q^{n}p^{m})\nonumber\\
&=&\{ q^{n}p^{m}, f(q,p) \}^{(-s)}_{MB}
\end{eqnarray}
where $f$ is an arbitrary c-number function and the s-parametrized
star product $\star _{(-s)}$ is defined to be
\begin{equation}
\star_{(-s)}= \exp \frac {1}{2}i\hbar
[(1-s)\partial^{L}_{p}\partial^{R}_{q}-
(1+s)\partial^{L}_{q}\partial^{R}_{p}] 
\end{equation}
$\partial^{L}$ and $\partial^{R}$ denote partial derivatives
acting to the left (L) and to the right (R), respectively.
We have in particular
$\exp(-i\hbar\partial^{L}_{q}\partial^{R}_{p})$ for $s=1$, 
$\exp(\frac{1}{2}i\hbar(\partial^{L}_{p}\partial^{R}_{q}-
\partial^{L}_{q}\partial^{R}_{p}))$ for $ s=0$ and
$\exp(i\hbar\partial^{L}_{p}\partial^{R}_{q})$ for $s=-1$. 
Thus the above definition unifies the different expressions  given 
in the literature for the star product
and Moyal brackets  and generalizes them for an arbitrary s-ordering. 
The s-Moyal brackets of two arbitrary functions 
can also be written as 
\begin{eqnarray}
\{f_{1},f_{2}\}^{(-s)}_{MB}=2i f_{1}[\exp-\frac{1}{2}i\hbar s 
(\partial^{L}_{p}\partial^{R}_{q}+\partial^{L}_{q}\partial^{R}_{p})]
\sin[\frac{1}{2}\hbar(\partial^{L}_{p}\partial^{R}_{q}-
\partial^{L}_{q}\partial^{R}_{p})]f_{2}
\end{eqnarray}
which reduces to the well known Moyal form when $s=0$.

Writing out
the first three terms explicitly, the expansion of the s-Moyal brackets is 
\begin{eqnarray}
\{f_{1},f_{2}\}^{(-s)}_{MB}=
i\hbar \{f_{1},f_{2}\}_{PB}+\frac{1}{2!}(i\hbar/2)^{2}4s
[(\partial^{2}_{q}f_{1})(\partial^{2}_{p}f_{2})-
(f_{1} \leftrightarrow f_{2})]\nonumber\\
+\frac{1}{3!}(i\hbar/2)^{3}
\{[(1-s)^{3}+(1+s)^{3}]
(\partial^{3}_{p}f_{1})(\partial^{3}_{q}f_{2})- \nonumber\\
6(1-s^{2})(\partial_{q}
\partial^{2}_{p}f_{1})(\partial_{p}\partial^{2}_{q}f_{2})-
(f_{1} \leftrightarrow f_{2}) \}+\dots
\end{eqnarray} 
where PB denotes the Poisson brackets.
This formula generalizes to arbitrary values of $s$ 
the expansions for some discrete values of $s$ that  previously appeared
in the literature. In particular we would like to note that in the case of
Wigner quantization $(s=0)$, the leading order correction to the PB is 
proportional to $\hbar^{2}$, while in all the other cases $(s \neq 0)$ 
the leading term is proportional to $\hbar$.

Taking $\hat{F}$ to be $\hat{t}^{(s)}_{kl}$ in Eq.(89) and using (91)
we obtain
\begin{eqnarray}
\Gamma^{(s)}_{nm}(-s)(q^{k}p^{l})&=&
Tr\{[\hat{t}^{(s)}_{kl},\hat{t}^{(s)}_{nm}]\hat{\Delta}_{qp}(-s)\}\nonumber\\
&=&\{q^{n}p^{m},q^{k}p^{l}\}^{(-s)}_{MB}
\end{eqnarray}
The last equality sets up a Lie algebra isomorphism between the quantum
$W_{\infty}$, that is the algebra generated by s-ordered products
under the usual Lie bracket action, and the algebra generated by the
monomials  $q^{n}p^{m}$ for $n,m \geq 0$
under the s-MB action.
Since $(i\hbar)^{-1}\{,\}^{(-s)}_{MB} \rightarrow \{,\}_{PB}$  as
$\hbar \rightarrow 0$, the essence of the full quantum $W_{\infty}$
can be captured on the classical phase space by simply deforming
the Poisson brackets to s-Moyal brackets. On the other hand we have another 
infinite algebra generated by
the operators $\Gamma^{(r)}_{nm}(s); n,m \geq 0$ indexed by two
ordering parameters $r,s$ 
and built up by the product of the Bopp operators,
that are concretely realized in the tangent
space of $\bf {R}^{2}$. This is the algebra that we referred to as
the classical $W_{\infty}$ in the introduction. As is seen
from (95), or more readily from
\begin{eqnarray}
[\Gamma^{(r)}_{nm}(s),\Gamma^{(r)}_{kl}(s)]\hat{\Delta}_{qp}(s)=
-[[\hat{t}^{(r)}_{nm},\hat{t}^{(r)}_{kl}],\hat{\Delta}_{qp}(s)]
\end{eqnarray}
the above mentioned isomorphic quantum $W_{\infty}$ algebras are
central extensions of this classical $W_{\infty}$. The vanishing
of the right hand side of (95) (or (96)) requires by the completeness of the
basis, that $\hat{t}^{(r)}_{nm}$ (or
$[\hat{t}^{(r)}_{nm},\hat{t}^{(r)}_{kl}]$) has to be proportional
to $\hat{I}$, while the vanishing of the left hand side requires
$\Gamma^{(r)}_{nm}$( or $[\Gamma^{(r)}_{nm}(s),\Gamma^{r}_{kl}(s)]$)
to be zero. Note that, as is apparent from Eq.(96), there is an
overall sign difference between the structure constants of the
classical and quantum $W_{\infty}$ algebras. This can be remedied
by a simple redefinition of the generators. Thus, the group generated by
the quantum $W_{\infty}$ provides a projective representation
of the classical $W_{\infty}$.

On the other hand, it is known that \cite{Pope} the space
of monomials $t_{nm}\equiv q^{n}p^{m}; n,m \geq 0$ form the
Lie algebra of canonical diffeomorphisms of a phase space, that is  
topologically equivalent to $\bf {R}^{2}$, under the usual Poisson brackets.
This algebra is known as $w_{\infty}$, or since the area element
and the symplectic form coincides in two dimensions, as the algebra
of area preserving diffeomorphisms $Diff_{A}\bf {R}^{2}$. The
$W_{\infty}$ algebras discussed above are the quantum (or,$\hbar$)
deformation of this classical $w_{\infty}$. The one called
the quantum $W_{\infty}$ provides
an implementation of the general canonical transformations
at the quantum level.
\section{CONCLUSION}
We have developed a derivative based approach to the
WWGM-quantization as an alternative to the integral based
conventional one. This enabled us  in particular to obtain some
fundamental associations in a unified way
easily and to derive an explicit
formula for the s-ordered products that led to further 
observations. It is argued that this formula can also be used
for any pair of operators. In the case of operators belonging to 
a nilpotent
algebra such as the Heisenberg-Weyl algebra, 
seemingly different but equivalent
expressions of a given ordered product can be obtained.

In a given association the primary
issue  is to determine  how a
member of the association 
transforms when the other one 
is transformed in a well defined way.
We have explicitly shown that the WWGM-quantization
in its most general form has a $W_{\infty}$-covariance
which includes the known metaplectic covariance as a
subset. Eq.(90) contains  for $n , m \leq 2$ the
generators of the metaplectic algebra  $Isp(2)$
in the classical phase space and its central extension
in $\cal{H}$. Moreover, we emphasize that like the
metaplectic covariance, the $W_{\infty}$-covariance 
we had shown is a
genuine property of the  complete operator bases used. An important
group theoretical outcome of this construction is that we
have obtained a projective representation of the classical
$W_{\infty}$ realized in the tangent space of the related
phase space.

$W_{\infty}$ algebras are currently the subject
of active investigations in two dimensional gravity \cite{Pope},
conformal field theories \cite{Bouwknegt} and in connection with quantum
Hall effect in condensed matter physics
(see \cite{Cappelli} and the references therein).
For example, the notion
of incompressibility which plays a fundamental role in
the theoretical understanding of  quantum Hall effect
has been related to the existence of the $W_{\infty}$
symmetry. These exciting developments suggest that the 
structure of the Landau
levels, or more generally the quantum Hall effect could also be
investigated in the framework of the WWGM-quantization.
We plan to take up a systematic
study of these problems in our forthcoming papers.

\vskip 4mm
\noindent {\bf Acknowledgments}
\vskip 2mm
A. Ver\c{c}in wishes to thank Middle East Technical University
for the warm hospitality. Our
special thanks are due to  B. S. Kandemir
for his help in preparing the Revtex file of the manuscript, and
to T. Altanhan for useful conversations. 
This work is supported in part by the Scientific and
Technical Research Council of Turkey (T\"{U}B\.{I}TAK)
and the Turkish Academy of Sciences (T\"{U}BA).


\begin{references}
\bibitem{Weyl} H. Weyl, {\bf The Theory of Groups and Quantum
Mechanics} (Dover, New York, 1931). \\
E. P. Wigner, Phys.Rev. {\bf 40} (1932) 749. \\
H. J. Groenewold, Physica {\bf 12} (1946) 405. \\
J. E.  Moyal, Proc. Camb. Phil. Soc. {\bf 45} (1949) 99. \\
\bibitem{Kubo}R. Kubo, J.Phys.Soc.Japan, {\bf 19} (1964) 2127. \\
K. \.{I}mre, E. \"{O}zizmir, M. Rosenbaum, and P. F. Zweifel, J.
Math. Phys. {\bf 8} (1967) 1097. \\
A. Alastuey, and B. Jancovici, Physica {\bf 97A} (1979) 349,
and {\bf 102A} (1980) 327. \\
\bibitem{Klauder}J. R. Klauder, and E. C. G. Sudarshan,{\bf  Fundamentals
of Quantum Optics} (Benjamin, New York, 1968). \\
J. R. Klauder, and E. S. Skagerstam,{\bf  Coherent States: Applications
in Physics and Mathematical Physics} (World
Scientific, Singapore,1985). \\
A. M. Perelomov, {\bf Generalized Coherent States and
Their Applications} (Springer, Berlin, 1986). \\
W. M. Zhang, D. H. Feng, and R. Gilmore, Rev. Mod.
Phys. {\bf 62} (1990) 867. \\
\bibitem{Carruthers}P. Carruthers, and F.
Zachariasen, Rev. Mod. Phys. {\bf 55} (1983) 245. \\
\bibitem{Takahashi}K. Takahashi, Prog. Theoret.
Phys. Suppl. {\bf 98} (1989) 109. \\
\bibitem{Lee} H. W. Lee, Phys. Rep. {\bf 259} (1994) 147. \\
\bibitem{Folland}G. B. Folland, {\bf Harmonic Analysis
in Phase space} (Princeton Univ. Press, Princeton, NJ, 1989).
\bibitem{Kirillov}A. A. Kirillov, {\bf Elements of the Theory of
Representations} (Springer-Verlag, Berlin, 1976).
\bibitem{Bayen}F. Bayen, M. Flato, C. Fronsdal, A.
Lichnerowicz, and D. Sternheimer, Ann. Phys. {\bf 111} (1978) 61, and 111. \\
\bibitem{Cahill} K. E. Cahill, and R. J. Glauber,
Phys. Rev. {\bf 177} (1969) 1857, and 1882. \\
\bibitem{Agarwal}G. S. Agarwal, and E. Wolf,
Phys. Rev. {\bf D2} (1970) 2161, 2187, and 2206.\\
\bibitem{Dunne}G. Dunne, J. Phys. A: Math. Gen. {\bf 21} (1988) 2321. \\
\bibitem{Littlejohn}R. G. Littlejohn, Phys. Rep. {\bf 138} (1986) 193. \\
V. Guillemin, and S. Sternberg, {\bf Symplectic Techniques
in Physics} (Cambridge University, Cambridge, 1984). \\
\bibitem{Han} D. Han, Y. S. Kim, and M. E. Noz,
Phys. Rev. {\bf A37} (1988) 807. \\
 Hui Li,  Phys. Lett. {\bf A188} (1994) 107,
 and {\bf A190} (1994) 370. \\
\bibitem{Osborn}T. A. Osborn, and F. H. Molahn,
Ann. Phys. {\bf 241} (1995) 79. \\
\bibitem{Pope}C. N. Pope, X. Shen, and L. J.
Romans, Nucl. Phys. {\bf B339} (1990) 191. \\
I. Bakas, Phys. Lett. {\bf B228} (1989) 57. \\
E. Bergshoeff, P. S. Howe, C. N. Pope, E. Sezgin, X.
Shen, and K. S. Stelle, Nucl. Phys. {\bf B.363} (1991) 163.\\
\bibitem{Balazs} N. L. Balazs, and B. K.
Jennings, Phys. Rep. {\bf 104} (1984) 347. \\
\bibitem{Hillery}M. Hillery, R. F. O'Connell, M. O.
Scully, and E. P. Wigner, Phys. Rep. {\bf 106} (1984) 121. \\
S. R. De Groot, and L. G. Suttorp, {\bf Foundation
of Electrodynamics} (North-Holland, Amsterdam, 1972) Ch.VI. \\
\bibitem{Bender}C. M.  Bender, and G. Dunne, Phys.
Rev. {\bf D40} (1989) 3504. \\
\bibitem{Gelfand}I. M. Gelfand, and D. B. Fairlie, Commun.
Math. Phys. {\bf 136} (1991) 487. \\
\bibitem{Bopp}F. Bopp, Ann. Inst. H. Poincar\'{e} {\bf 15} (1956) 81. \\
\bibitem{vercin} A. Ver\c{c}in, {\it Metaplectic covariance of the Weyl-Wigner-Groenewold-Moyal quantization and beyond}, preprint (1997)  
\bibitem{Bouwknegt}P. Bouwknegt, and K. Schoutens,
Phys. Rep. {\bf 223} (1993) 184.\\
\bibitem{Cappelli}A. Cappelli, C. A. Trungenberger, and G. R. Zemba,
Nucl. Phys. {\bf B448} [FS] (1995) 470. \\
D. Karabali, Nucl. Phys. {\bf B428} [FS] (1994) 531. \\
\end{references}
\end{document}